\documentclass[reprint,amsmath,amssymb, aps, prc,preprintnumbers]{revtex4-1}
\usepackage{graphicx}
\usepackage{dcolumn}
\usepackage{bm}
\usepackage{pgfplotstable}
\usepackage{floatpag}
\usepackage{tabularx}
\usepackage{hyperref}
\usepackage[subpreambles=true]{standalone}
\usepackage[caption=false]{subfig}
\usepackage{pgfplots}
\usepackage{longtable}
\usepackage{booktabs}
\usepackage{flafter}
\usepackage{pgf}
\usepackage[utf8]{inputenc}

\newcommand{\gev}{\,\mathrm{GeV}}

\begin{document}
\title{Charge symmetry violation in the determination
  of strangeness form factors}

\preprint{ADP-19-3/T1083}

\author{Ali Alkathiri}
\altaffiliation{On leave from: Physics Department, Taif
    University, Taif 26571, Saudi Arabia.}
\affiliation{CSSM, School of Physical Sciences, University of
  Adelaide, Adelaide SA 5005, Australia}
\author{Ross D.~Young, James M.~Zanotti}
\affiliation{CSSM, School of Physical Sciences, University of
  Adelaide, Adelaide SA 5005, Australia}

\begin{abstract}
  The strange quark contributions to the electromagnetic form
  factors of the proton are ideal quantities to study the role of
  hidden flavor in the properties of the proton.
  This has motivated intense experimental measurements of these form
  factors.
  A major remaining source of systematic uncertainty in these
  determinations is the assumption that charge symmetry violation
  (CSV) is negligible.
  We use recent theoretical determinations of the CSV form factors
  and reanalyse the available parity-violating electron scattering
  data, up to $Q^2\sim1$GeV$^2$.
  Our analysis considers systematic expansions of the strangeness
  electric and magnetic
  form factors of the proton.
  The results provide an update to the determination of strangeness
  over a range of $Q^2$ where, under certain assumptions about the
  effective axial form factor, an emergence of non-zero strangeness is
  revealed in the vicinity of $Q^2\sim 0.6\,{\rm GeV}^2$.
  Given the recent theoretical calculations, it is found that the
  current limits on CSV do not have a significant impact on the
  interpretation of the measurements and hence suggests an opportunity
  for a next generation of parity-violating measurements to more
  precisely map the distribution of strange quarks.
\end{abstract}

\maketitle

\section{Introduction}
The desire for a complete understanding of the electromagnetic
structure of the proton has led to significant efforts over the last
two decades to determine the individual quark flavor contributions to
the proton's electromagnetic form factors.
A significant challenge in this goal lies in determining the role
played by non-valence or ``hidden'' quark flavors whose contributions
arise only through fluctuations of the QCD vacuum.
Being the lightest sea-only quark, strange quarks are anticipated to
make the most significant contribution.
Through an extensive experimental program of parity-violating electron
scattering (PVES) \cite{35,37,45,46,147,39,42,148,149,150,43,44},
strange quarks have been tagged by measuring the neutral-current form
factors.
The isolation of strangeness relies on the assumption of good charge
symmetry, which has been one of the limiting factors in extending the
experimental program to greater precision.
In this work, we quantify the impact of charge symmetry violation
(CSV) on the extraction of strangeness from a global analysis of the
PVES measurements.

While earlier theoretical predictions of CSV in the proton's
electromagnetic form factors varied through several orders of
magnitude \cite{18,120,173, 318}, a recent lattice QCD calculation
\cite{151} has determined that CSV in the proton's electromagnetic
form factors is significantly smaller than earlier expectations.
Despite its importance for future measurements of parity violating
electron-proton scattering and their subsequent interpretation as
evidence of proton strangeness, the precise influence of this recent
CSV constraint has not been throughly quantified.
Hence, here we perform a global analysis of the full existing set of
parity-violating (PV) asymmetry data with and without the constraint of
CSV form factors.
To achieve this task, we consider PVES data, obtained from experiments
conducted with varying kinematics and targets, from
SAMPLE~\cite{35,37}, PVA4~\cite{45,46,147},
HAPPEX~\cite{39,42,148,149,150}, G0~\cite{43,44} and
Q$_\text{weak}$~\cite{73}.

This paper is organised as follows: In section~\ref{sec2}, we describe
the formalism of PVES, including PV asymmetries of the nucleon,
helium-4 and the deuteron.
Section~\ref{sec3} presents the parametrisation of strange quark form
factor, while section~\ref{sec4} is dedicated to a study of the CSV
effects on strangeness form factor extraction.
A brief summary is finally presented in section~\ref{sec5}.

\section{STRANGE FORM FACTORS AND PARITY-VIOLATING ELECTRON
  SCATTERING}
\label{sec2}
\begin{figure}[t]
\centering
\includegraphics[width=\columnwidth]{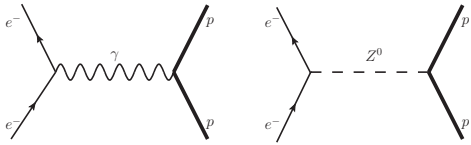}
\vspace{-28em}
\caption{Total leading order amplitude for electron-nucleon scattering
  is the sum of the leading order electromagnetic and neutral current
  amplitudes.}
\label{Fig120}
\end{figure}

Determining the strange electric and magnetic form factors
experimentally requires a process where the weak and electromagnetic
interactions interfere.
This is achieved through PVES experiments
\cite{Beck:1989tg,Mckeown:1989ir}, whose leading-order amplitudes are
shown in Fig.~\ref{Fig120}.

Under the assumption of charge symmetry, the PV asymmetry in polarised
$e$--$p$ scattering is given by~\cite{64}
\begin{equation}
\label{eq10}
A_{PV}^p = \Bigg[ \frac{-G_FQ^2}{4\sqrt{2}\pi\alpha} \Bigg] 
(A^p_V + A^p_s +A^p_A)\,.
\end{equation}
where in terms of the proton's electric $(G^{\gamma,p}_E)$ and
magnetic $(G^{\gamma,p}_M)$ Sach's form factors
\begin{align}
\label{eq17}
A^p_V =& (1 - 4 \sin^2\hat\theta_\text{w}) (1 + R^p_V)\\ \nonumber
&- (1 + R^n_V) \frac{\epsilon G^{\gamma,p}_E G^{\gamma,n}_E +
  \tau G^{\gamma,p}_M G^{\gamma,n}_M }
{\epsilon(G^{\gamma,p}_E)^2 + \tau(G^{\gamma,p}_M)^2}\,,
\end{align}
\begin{equation}
\label{eq18}
A^p_s = -(1+R^{(0)}_V) \frac{\epsilon G^{\gamma,p}_E G^{s}_E +
  \tau G^{\gamma,p}_M G^{s}_M }
{\epsilon(G^{\gamma,p}_E)^2 + \tau(G^{\gamma,p}_M)^2}\,,
\end{equation}
and
\begin{equation}
\label{eq19}
A^p_A = \frac{-\epsilon^{\prime} (1 - 4\sin^2\hat\theta_W)
  G^{\gamma,p}_M {\tilde{G}^{e,p}_A}}
{\epsilon(G^{\gamma,p}_E)^2 + \tau(G^{\gamma,p}_M)^2}.
\end{equation}
are the proton's vector form factor excluding strangeness $(A^p_V)$,
the proton's strangeness vector form factor $(A^p_s)$, and the
interference of the proton's magnetic vector and the axial vector form
factors $(A^p_A)$.

The electromagnetic form factors of the proton and neutron are denoted
by $G_{E,M}^\gamma$, the strangeness vector form factors $G_{E,M}^s$,
and the effective axial form factor $\tilde{G}_A^{p}$.
The kinematic variables, which depend on the four-momentum transfer
$Q^2=-q^2$ and the electron scattering angle $\theta$, are defined as
\begin{equation}
\label{eq5}
\tau=\frac{Q^2}{4M^2_p},
\end{equation}
\begin{equation}
\label{eq6}
\epsilon=\frac{1}{1+2(1+\tau)\tan^2\frac{\theta}{2}},
\end{equation}
and
\begin{equation}
\label{eq7}
\epsilon^{\prime}=\sqrt{\tau(1+\tau)(1-\epsilon^2)},
\end{equation}
where $M_p$, $\epsilon$ and $\epsilon^{\prime}$ are the proton's mass,
the virtual photon longitudinal polarization and the scattered energy,
respectively.
The standard model parameters: fine structure constant $\alpha$, Fermi
coupling $G_F=1.16638\times 10^{-5}\gev^2$, the weak mixing angle,
$\sin^2\hat{\theta}_W=0.23129$, in the $\overline{\text{MS}}$
renormalization scheme, and the standard model radiative corrections,
$R^p_V=-0.0513$ and $R^n_V=R^{(0)}_V=-0.0098$ are all obtained from
the PDG ~\cite{Tanabashi:2018oca}.

\subsection{Helium-4 and Deuteron PV Asymmetries}
The $^4$He nucleus is spin zero, parity even and isoscalar.
Elastic electron scattering from $^4$He is an isoscalar $0^+ \to 0^+$
transition and therefore allows no contributions from magnetic or
axial-vector currents.
Thus, the HAPPEX Collaboration has utilised a $^4$He target to
directly extract the strange electric form factor~\cite{39}.
Nuclear corrections are relevant for $^4$He and deuteron targets, more
details about which can be found in~\cite{190,171}.
 
The theoretical asymmetry based on the assumption that isospin mixing
can be neglected is written as~\cite{64}
\begin{align}
\label{eq20}
A^{He}_{PV} =& \Bigg[ \frac{G_FQ^2}{4\sqrt{2}\pi\alpha} \Bigg]
. \Bigg[ (1-4\sin^2\theta_\text{w}) (1+R^p_V) - (1+R^n_V)\\ \nonumber
&+ 2\frac{-\big(1+R^{(0)}_V\big)G^s_E}{G^p_E+G^n_E} \Bigg] .
\end{align}
Corrections associated with isospin violations will be considered in
Section~\ref{sec4}.

The parity-violating asymmetries measured in quasi-elastic scattering
from the deuteron have responses which involve the deuteron
wavefunctions. For this analysis, we use directly the theoretical
asymmetries as reported with original experimental measurements.
Early results from SAMPLE have had minor modifications made
to update for more recent radiative corrections and form factor
parameterisations, as described in Ref.~\cite{152}.

\subsection{$\gamma Z$-exchange corrections to PVES}
\label{sec2.3}
Leading electroweak corrections play a significant role in precision
measurements of the strangeness contribution to the nucleon form
factors \cite{Musolf:1990ts,64}.
In contrast to the formalism relevant to atomic parity violation
experiments \cite{Marciano:1982mm}, an energy-dependent correction
arising from the $\gamma Z$ box diagram was highlighted by Gorchtein
\& Horowitz \cite{Gorchtein:2008px}.
The size of this correction is particularly significant to the
standard model test by the Q-weak Experiment \cite{Androic:2018kni}.
Fortunately, the uncertainties arising from the underlying $\gamma Z$
interference structure functions can be reliably constrained
\cite{292}.
These results have been further supported by direct measurement
\cite{Wang:2014bba}.

The significance of the $\gamma Z$ box is somewhat less pronounced in
the determination of strangeness.
Nevertheless, for example, the correction makes about $\sim
\tfrac12$-sigma shift to the central value of the precise HAPPEX
proton point at $Q^2\sim0.1\gev^2$.
We incorporate the corrections reported by the constrained model of
Ref.~\cite{292}, updated with the improved constraints of quark-hadron
duality \cite{Hall:2015loa}, and a momentum dependence as proposed in
Ref.~\cite{74}.
For completeness, a table of values is included in
Appendix~\ref{app:gZ}.

\subsection{Theoretical asymmetries}
\label{sec2.4}
In this work, a set of all available PV asymmetry data up to $Q^2\sim
1 \,$GeV$^2$, as summarised in Table~\ref{tab1}, is analysed.
Such a combined analysis of the world PV data requires a consistent
treatment of the vector and axial form factors and radiative
corrections. The theoretical asymmetry used in this analysis is
written as
\begin{equation}
\label{eq22}
A_{Theory} = \eta_0 + \eta^p_A\tilde{G}^p_A + \eta^n_A\tilde{G}^n_A +
\eta_EG^s_E + \eta_MG^s_M,
\end{equation}
where the $\eta$'s, provided in Table~\ref{tab1}, are calculated using
the recent elastic form factor parameterisations of Ye et
al.~\cite{Ye:2017gyb} and current values for the standard model
radiative corrections \cite{Tanabashi:2018oca}.

In this analysis, since the entire contribution is to be fit to data,
we employ the effective axial form factors $\tilde{G}^{p,n}_A$ which
implicitly includes both the axial radiative and anapole corrections.
For these form factors, we employ a dipole form 
\begin{equation}
\label{eq29}
\tilde{G}^{p,n}_A = \tilde{g}^{p,n}_A\bigg( 1 + \frac{Q^2}{M_A^2}
\bigg)^{-2}\, ,
\end{equation}
with an axial dipole mass $M_A = 1.026$ GeV, determined from neutrino
scattering~\cite{183}, common to both proton and neutron form factors.
The normalisations $\tilde{g}^{p,n}$ are fit to the data, however
since the isoscalar combination is very poorly determined, we choose
to impose theoretical esimates based on an effective field theory
(EFT) with
vector-meson dominance (VMD) model to constrain this combination,
$(\tilde{g}^p_A+\tilde{g}^n_A)/2=-0.08\pm0.26$ \cite{165}.

\section{Analysis framework}
\label{sec3}

\subsection{Taylor expansion}
\label{sec3.1}
At low momentum transfers, a Taylor series expansion of the
electromagnetic form factors in $Q^2$ is sufficient and minimises the
model dependence.
Given the sparsity and precision of the available data---up to $\sim 1\gev^2$---we
avoid introducing a specific model by first attempting to parameterise the strange electric
and magnetic form factors $Q^2$ dependence by
\begin{align}
\label{eq:Taynlo}
G^s_E &= \rho_sQ^2 + \rho_s'Q^4\,,\nonumber\\
G^s_M &= \mu_s + \mu_s'Q^2\,.
\end{align}

\subsection{$z$-expansion}
\label{sec3.2}
{\em A priori}, one might not expect a Taylor expansion up to $\sim
1\,{\rm GeV}^2$ to be satisfactory.
To provide an alternative functional form to the Taylor expansion, we
also consider the $z$-expansion which offers improved convergence
based on the analytic properties of the form factors
\cite{Cutkosky:1969iv,188,187}.
We describe the momentum dependence of the strange form factors using the
$z$-expansion, also to second (nontrivial) order:
\begin{align}
\label{eq:znlo}
G^s_E = \rho_{s,z} z + \rho_{s,z}' z^2,\nonumber\\
G^s_M = \mu_s + \mu_{s,z}' z, 
\end{align}
where
\begin{equation}
\label{eq36}
z = \frac{\sqrt{t_{cut}+Q^2} - \sqrt{t_{cut}}}
{\sqrt{t_{cut} + Q^2} + \sqrt{t_{cut}}}\,.
\end{equation}
In our fits, we use choose $t_{cut}=(2 m_{K})^2$, with the kaon mass
$m_K=0.494$ GeV.
In the absence of isospin violation, the cut formally starts at
$9m_\pi^2$, but we assume that the strangeness contribution to the
3-pion state can be neglected.
We note that, with the current experimental precision, there isn't any
significant sensitivity to the value of $t_{cut}$.
To more easily facilitate the comparison with the two expansion forms,
we report the simple Taylor expansion coefficients for each case.
That is, for the $z$ fits, we translate the expansions back in the
leading Taylor form, e.g.~$\rho_s=dG^s_E/dQ^2|_{Q^2=0}$.

\subsection{Charge symmetric results}
\label{sec3.3}
Here we summarise the fit results under the assumption of exact charge
symmetry.
This provides a baseline with which to explore the implications of
charge symmetry violation in the following section.

In this work we perform a global fit at leading order (LO) and at next
leading order (NLO) of the strangeness form factor.
Thus, the fitting procedure at LO considers four parameters,
$\tilde{g}^p_A$, $\tilde{g}^n_A$, $\mu_s$ and $\rho_s$, while fitting
at NLO considers an additional two parameters, $\mu'_s$ and $\rho'_s$.

The $\chi^2$ is calculated as
\begin{equation}
\label{eq30}
\chi^2 = \sum_i \sum_j (m_i - t_i)(V)^{-1}_{ij}(m_j - t_j)\,,
\end{equation}
where $m$ and $t$ denote the measurement and theory asymmetry
respectively.
The indicies $i$ and $j$ run over the data ensemble.
The matrix $V$ represents the covariance error matrix defined as
\begin{equation}
\label{eq31}
(V)_{ij} = (\sigma^{uncor}_i)^2 \delta_{ij} + \sigma^{cor}_i
\sigma^{cor}_j\,,
\end{equation}
where $\sigma^{uncor}_i$ and $\sigma^{cor}_i$ are uncorrelated and
correlated uncertainties of the $i^{\text{th}}$-measurement,
respectively.
We note that the correlated uncertainties are only relevant for the G0
experiment, where the forward \cite{43} and backward \cite{44} are
treated as mutually independent.
The goodness of fit is estimated from the reduced $\chi^2$ as
\begin{equation}
\label{eq32}
\chi^2_{red} = \chi^2/\text{d.o.f}\,,
\end{equation}
with 33 and 31 degrees of freedom (d.o.f) for the LO and NLO fits,
respectively.
Note that with the isoscalar axial ``charge'' constrained, as
described above, there are effectively 3(5) fit parameters in the
LO(NLO) fits.
\begin{table*}
  \caption{\label{tab2}
    The parameter values and $\chi^2$ obtained from previous
    PVES global fits~\cite{152,160,167,Liu:2007yi} and the current global
    analysis at LO for both Taylor and $z$-expansion form factor fits
    without constraints from CSV.}
\begin{ruledtabular}
  \begin{tabular}{l|ccc}
                           & {$\rho_s$ [GeV$^{-2}$]}    & {$\mu_s$}         & {$\chi^2_{red}$}\\
    \hline
    YRCT(2006) \cite{152}       &  $-0.06\pm0.41$           &  0.12$\pm$0.55    & 1.3 \\
    YRCT(2007) \cite{167}       &  $0.02\pm0.18$            & $-0.01\pm0.25$    & ---\\
    LMR(2007) \cite{Liu:2007yi} & $-0.08\pm0.16$            & $0.29\pm0.21$     & 1.3\\
    GCD(2014) \cite{160}        &  0.26$\pm$0.16            &  $-0.26\pm0.26$   & 1.3 \\
    \hline
    Taylor          &  0.15$\pm$0.04        &  -0.12$\pm$0.04  & 1.1 \\
    $z$-exp.        &  0.18$\pm$0.05        &  -0.10$\pm$0.04  & 1.1
  \end{tabular}
\end{ruledtabular}
\end{table*}
\begin{table*}
  \caption{\label{tab3}
    The NLO parameters values and $\chi^2$ obtained from a previous
    global fit~\cite{152}, where $Q^2<0.3$ GeV$^2$, and the current global
    analysis at NLO for both Taylor and $z$-expansion form factor fits
    without constraints from CSV.}
\begin{ruledtabular}
  \begin{tabular}{l|ccccc}
                       & $\rho_s$ [GeV$^{-2}$] & $\rho_s'$ [GeV$^{-4}$] & $\mu_s$        & $\mu_s'$ [GeV$^{-2}$] & $\chi^2_{red}$\\
 \hline
 YRCT(2006) \cite{152} &  $-0.03\pm0.63$       & $-1.5\pm 5.8$          & $0.37\pm0.79$  & $0.7\pm 6.8$          &  1.4 \\
 \hline
 Taylor                &  0.07$\pm$0.14        & 0.14$\pm$0.22          & -0.05$\pm$0.15 & -0.11$\pm$0.23        & 1.23 \\
 $z$-exp.              &  0.08$\pm$0.17        & 0.19$\pm$0.37          & -0.09$\pm$0.14 & -0.06$\pm$0.29        & 1.26
  \end{tabular}
\end{ruledtabular}
\end{table*} 

We report the leading-order fit results in Table \ref{tab2}, with
comparisons against previous work.
The results are compatible with earlier work, though with
significantly reduced uncertainty.
This is due to both an updated list of measurements and the inclusion
of the full range of $Q^2$ points in the fit.
No appreciable difference is seen between the simple Taylor expansion
and the $z$-expansion.

While the fit quality is reasonable, these simple leading-order fits
are certainly anticipated to be too simple to describe these form
factors over the full range $0\le Q^2\lesssim 1.0\gev^2$.
As a result, the statistical uncertainties displayed are not
representative of the current knowledge of the strange form factors.
We hence allow for more variation in the $Q^2$ dependence by extending
the fits to next-leading order, Eqs.~(\ref{eq:Taynlo}) and
(\ref{eq:znlo}).
Results are shown in Table~\ref{tab3}.
Curiously the additional fit parameters are unable to make significant
improvement to the $\chi^2$ and the reduced $\chi^2$ very marginally
increases for the NLO fit.

Although the data do not support any structure offered by the NLO
fits, we prefer the results at this order as being better
representative of the uncertainties of the strangeness form factors,
while offering some degree of smoothing of the underlying data.
We note that given the clustering of the underlying data set, the
separation of the electric and magnetic strange form factors are only
most reliable at the discrete momentum transfers near $Q^2\sim 0.1, 0.2$ and
$0.6\gev^2$.
As such, the NLO fits are roughly fitting 3 data points with 2
parameters for each form factor.
Attempting further higher order will just amount to over-fitting the
statistical fluctuations of the data set.

\begin{figure}[t]
\centering
\includegraphics[width=\columnwidth]{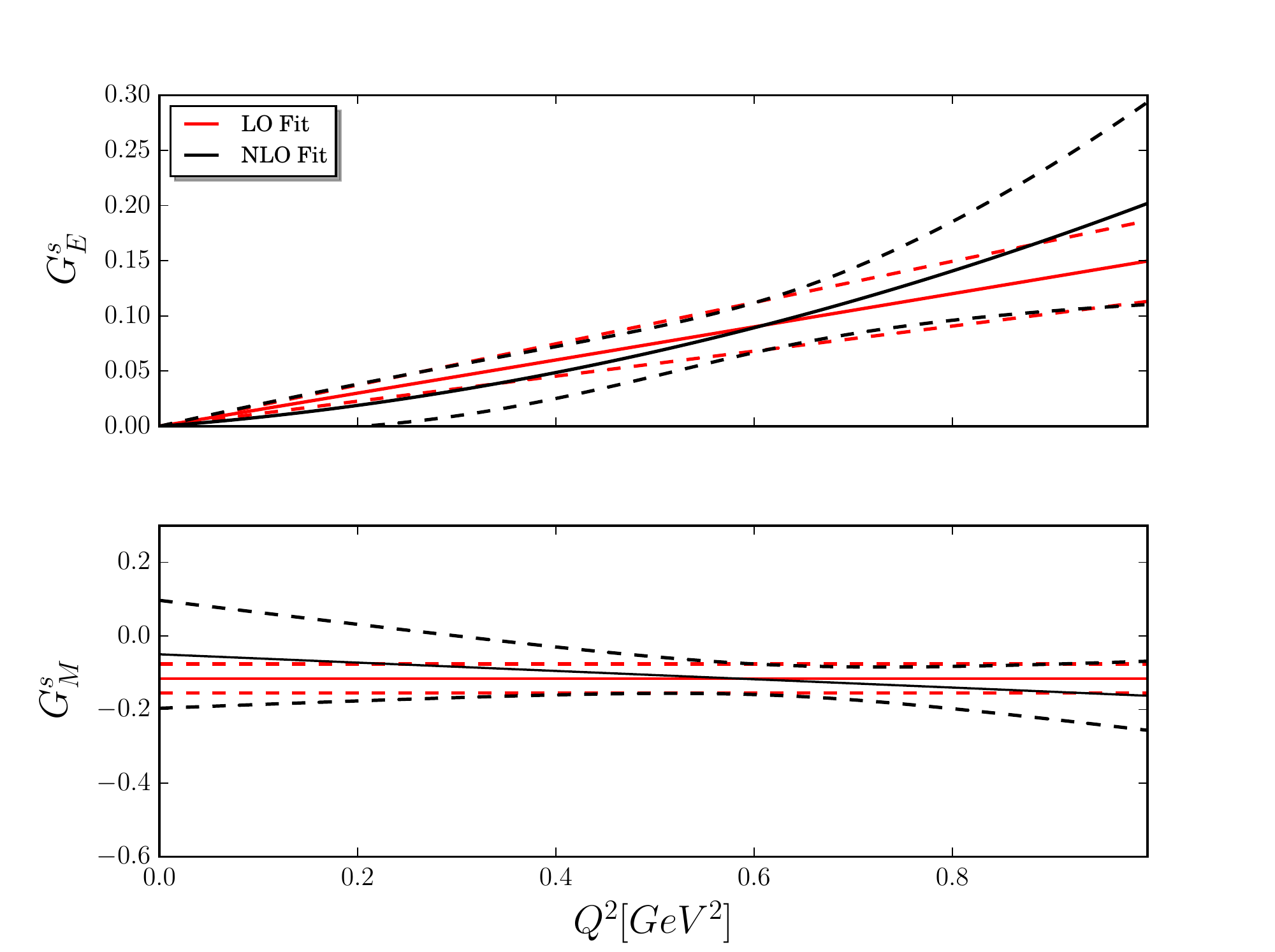}
\caption{The extracted strange electric and magnetic form factors from a
  global fit up to $Q^2\sim 1$ GeV$^2$ using the Taylor expansions in
  Eq.~(\ref{eq:Taynlo}).
  The red (black) solid curve shows the LO (NLO) fit and bound
  shown by the dotted curves.}
\label{Fig11}
\end{figure}

\begin{figure}[t]
\centering
\scalebox{0.45}{\input{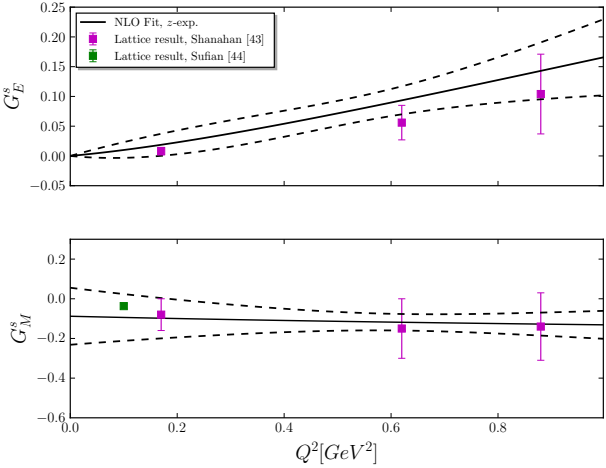}}
\caption{The extracted strange electric and magnetic form factors from
  global fit up to $Q^2\sim 1$ GeV$^2$ with using the NLO $z$-expansion
  in Eq.~(\ref{eq:znlo}).
  A comparison with recent lattice QCD results is shown where the
  green square (errors bars smaller than the symbol) corresponds to
  the result of $G^s_M$($Q^2=0.1$ GeV$^2$)~\cite{191} and the magenta
  squares are $G^s_M$ and $G^s_E$ at 
  $Q^2=0.17,\ 0.62$ and 0.88 GeV$^2$~\cite{146}.}
\label{Fig100}
\end{figure}

The comparison between the leading and next-to-leading order fits for
the Taylor expansion are shown in Figure~\ref{Fig11}.
The corresponding $z$-expansion results are very similar.
In Fig.~\ref{Fig100} we show the NLO $z$-expansion parameterisation of the
separated electric and magnetic form factors and compare with recent lattice
QCD results \cite{146,191}.
Here we observe excellent agreement between our strangeness determination based on
PVES data and lattice QCD results over the full
$Q^2$ range.
These are also compatible with earlier lattice \cite{184} and
lattice-constrained \cite{31,51,126} results.
Interestingly, we note that the experimental results are showing some
support for a non-vanishing strangeness electric form factor in the
vicinity of $Q^2\sim 0.6\gev^2$.
Given the lack of sensitivity to the choice of functional form, we
adopt the $z$-expansion at NLO as our preferred fit for the following
discussions.

\begin{figure}[!htp]
\centering
\includegraphics[width=\columnwidth]{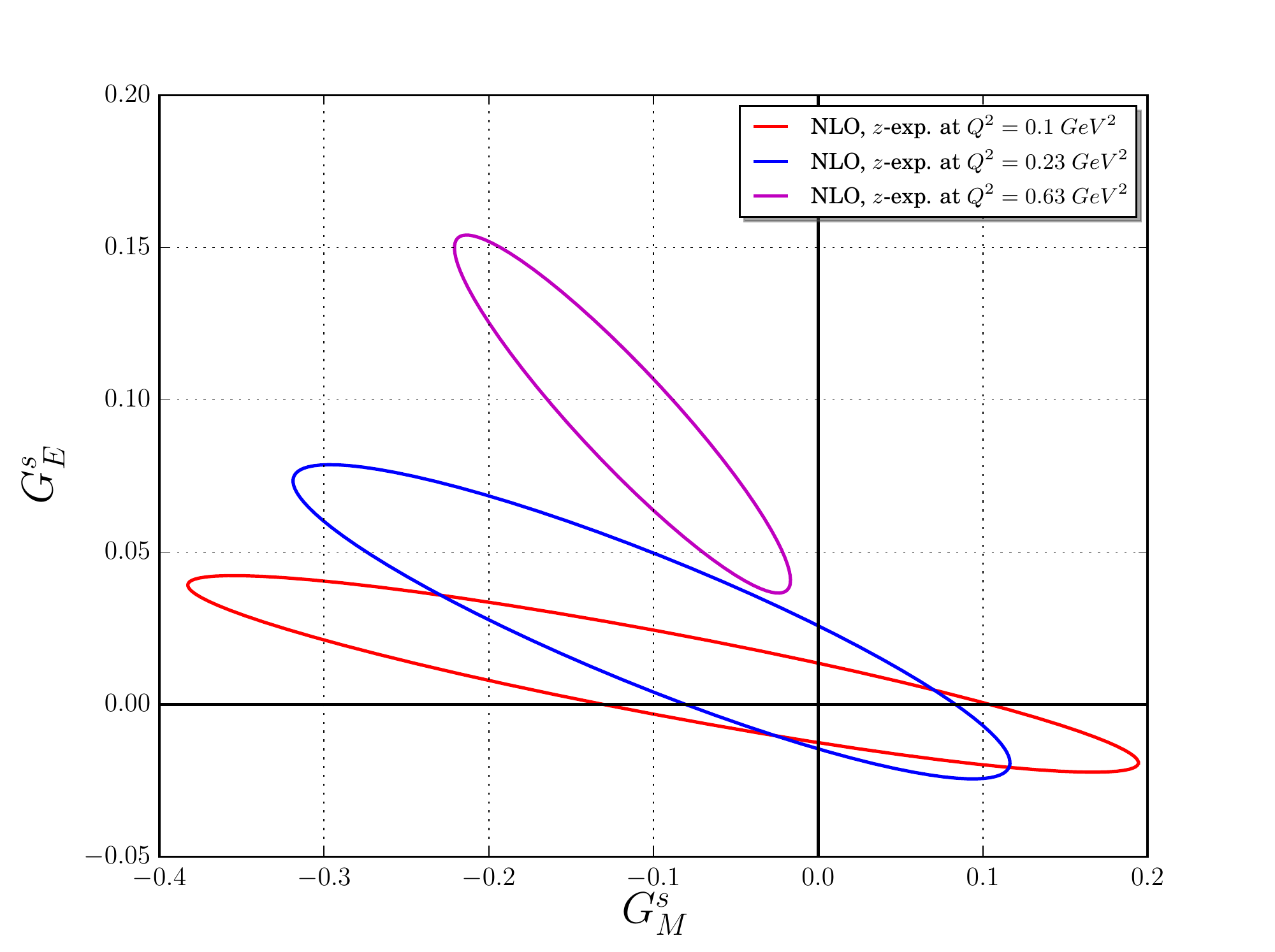}
\caption{95\% confidence level ellipses for the electric and magnetic
  strangeness form factors using the NLO $z$-expansion in
  Eq.~(\ref{eq:znlo}) for three $Q^2$ values 0.1, 0.23, 0.63 GeV$^2$.}
\label{Fig102}
\end{figure}

The preceeding discussion has focussed on the separation of the
electric and magnetic strangeness form factors.
Given the high degree of correlation in the measurements, it is instructive to
display the joint confidence intervals.
Fig.~\ref{Fig102} displays the 95$\%$ confidence level ellipses for
the different values of $Q^2=0.1, 0.23$ and $0.63\gev^2$ for the NLO
$z$-expansion fit.
At the low $Q^2$ values, we observe that the strangeness form factors
are compatible with zero at the 95\% CL, with a marginal preference for positive values of
strange electric form factor and negative values of the magnetisation
--- as seen earlier in Refs.~\cite{167,169}.

At $Q^2=0.63\gev^2$, there appears a clear signal for nonzero
strangeness, with a negative $G_M^s$ and positive $G_E^s$.
In contrast to earlier work that has suggested vanishing strangeness
at this $Q^2$ \cite{44,150}, the dominant difference in the present work
is the treatment of the axial/anapole form factor.
As described, the isoscalar combination is constrained by the EFT and
VMD estimate of Zhu~et~al.~\cite{165}, while the isovector combination
is determined by the data.
The best fit---for $z$-expansion at NLO---results in
$\tilde{g}_A^{p}=-0.67\pm 0.25$, which is less negative than the
zero-anapole approximation.
As a consequence, the data-driven fit drives the back-angle G0 results
to be more consistent with a negative $G_M^s$.
Under these assumptions for the effective axial form factor, we see
$G_E^s\sim 0.1$, which---with the strange charge factor included---is
on the order of 10\% of the proton electric form factor at this
momentum transfer.
Given the smallness of the strange form factors determined in direct
lattice calculations \cite{184,191}, the result seen here suggests a
future investigation of the anapole ``charge'' and $Q^2$ dependence is
warranted.

\section{Sensitivity to Charge symmetry violation}
\label{sec4}
\subsection{CSV in asymmetries}
In this section we study the effects of charge symmetry violation
on our results, i.e. we no longer have a relationship between
the individual quark flavour contributions to the proton and neutron
form factors
\begin{align}\nonumber
G^{p,u}_{E,M}&\neq G^{n,d}_{E,M},\\ \nonumber
G^{p,d}_{E,M}&\neq G^{n,u}_{E,M}.
\end{align}
We follow standard notation and define the CSV form factors as
\begin{equation}
G^{CSV}_{E,M} = \frac{2}{3} (G^{p,d}_{E,M} - G^{n,u}_{E,M})
- \frac{1}{3} (G^{p,u}_{E,M} - G^{n,d}_{E,M})\,.
\end{equation}
In order to explore the impact of CSV, we need to modify the neutral
weak from factors to explicitly include a CSV term
\begin{align}
\label{eq140}\nonumber
G_{E,M}^{Z,p} =& (1 - 4\sin^2 \hat{\theta}_W) (1 + R^p_V)
G^{\gamma,p}_{E,M}\\ 
&- (1 + R^n_V) G^{\gamma,n}_{E,M} -
(1 + R^{(0)}_V) G^s_{E,M}\\
&- (1 + R^{n}_V) G^{CSV}_{E,M},\nonumber
\end{align}
where the $Q^2$-dependence of each form factor has been dropped for
clarity.
The CSV form factor can be expressed as a simple Taylor expansion in
$Q^2$
\begin{equation}
\label{eq35}
G^{CSV}_{E,M}(Q^2) = G^{CSV}_{E,M}(0) - \rho^{CSV}_{E,M}Q^2 +
\mathcal{O}(Q^4),
\end{equation}
with $G^{CSV}_{E}$ set to zero at $Q^2=0$ due to charge conservation.

Regarding the theoretical asymmetry given in Eq.~\ref{eq22}, we note
that $\eta_0$ will receive a correction due to the CSV form factor.
Hence
\begin{equation}
\label{eq135}
A_{Theory} = \eta^{CSV}_0 + \eta^p_A \tilde{G}^p_A +
\eta^n_A \tilde{G}^n_A + \eta_E G^s_E + \eta_M G^s_M\,,
\end{equation}
where
$\eta^{CSV}_0 = \eta_0 + \eta^{CSV}_{E} G^{CSV}_{E} + \eta^{CSV}_{M}
G^{CSV}_{M}$,
with
\begin{equation}
\label{eq33}
\eta^N_{_{CSV,E}} = \Big[ \frac{G_FQ^2}{4\sqrt{2}\pi\alpha} \Big] . 
\Big[ \frac{(1 + R^n_V) \epsilon G^{\gamma,N}_E}
  {\epsilon (G^{\gamma,N}_E)^2 + \tau (G^{\gamma,N}_M)^2} \Big]\,,
\end{equation}
\begin{equation}
\label{eq34}
\eta^N_{_{CSV,M}} = \Big[ \frac{G_FQ^2}{4\sqrt{2}\pi\alpha} \Big] .
\Big[ \frac{(1 + R^n_V) \tau G^{\gamma,N}_M}
  {\epsilon (G^{\gamma,N}_E)^2 + \tau (G^{\gamma,N}_M)^2} \Big]\,.
\end{equation}

In the case of the PV asymmetry of $^4$He, we should consider nuclear
CSV, which we denote $F^{CSV}$, in addition to CSV at nucleon level,
$G^{CSV}_E$.
In this case, $\eta^{CSV}_0$ can therefore be written as
\begin{equation}
\label{eq136}
\eta^{CSV}_0 = \eta_0-2 F^{CSV} +
4\frac{\left(1 + R^{n}_V \right) G^{CSV}_E}{G^p_E + G^n_E}\,,
\end{equation}
where in the notation of~\cite{170}, $F^{CSV} \equiv
F^{(1)}(q)/F^{(0)}(q)=-0.00157$ is used to calculate $\eta^{CSV}_0$
for the theoretical PV asymmetry of $^4$He at $Q^2=0.077$ and 0.091
GeV$^2$.

\subsection{CSV Theoretical Works}
\label{sec4.1}
To include effects of charge symmetry violation in our determination
of the strangeness form factors, we consider three different
calculations of the CSV form factors.
The first work we consider is from Kubis and Lewis~\cite{18}, denoted
by ``K$\&$L CSV''.
They used effective field theory, supplemented with resonance
saturation to estimate the relevant contact term --- where the CSV is
largely driven by $\rho-\omega$ mixing.
To accomplish this, they employ a large $\omega$-nucleon coupling
constant $g_\omega\sim42$ taken from dispersion analysis.
Combining this estimate with calculations in HB$\chi$PT and infrared
regularised baryon chiral perturbation theory, K$\&$L predicted a CSV
magnetic moment contribution $G^{CSV}_{M}(0)\equiv k^{u,d} = 0.025 \pm
0.020$, which includes an uncertainty arising from the resonance
parameter.
For the CSV slope parameters, K$\&$L found
$\rho^{CSV}_{M}=-0.08\pm0.06$~GeV$^{-2}$ and
$\rho^{CSV}_{E}=-0.055\pm0.015$~GeV$^{-2}$.
We take these values as our first estimate of the CSV form factors.

The second theoretical calculation of CSV we consider is from Wagman
and Miller~\cite{120}, denoted by ``W$\&$M CSV''.
In their work, they used relativistic chiral perturbation theory with
a more realistic $\omega$-nucleon coupling i.e. $g_\omega\sim 10$.
That study reported values of $G^{CSV}_{M}(0)=0.012\pm0.003$,
$\rho^{CSV}_{M}=0.015\pm0.010$ GeV$^{-2}$ and
$\rho^{CSV}_{E}=-0.018\pm0.003$ GeV$^{-2}$.

The third determination of the CSV form factor that we employ is based
on an analysis of lattice QCD results~\cite{151}, that we refer to
as ``Lattice CSV''.
The lattice study found significantly smaller values of the magnetic
and electric CSV form factors compared to the previous two estimates.
To study the effect of the CSV form factors obtained from lattice QCD,
we summarise the results of Ref.~\cite{151} by $G^{CSV}_{M}=0.0\pm0.001$ and
$\rho^{CSV}_{E}=0.0\pm0.001\gev^{-2}$.

\subsection{Strangeness with CSV}
\label{sec4.2}
\begin{figure*}[t]
\centering
\includegraphics[width=\columnwidth]{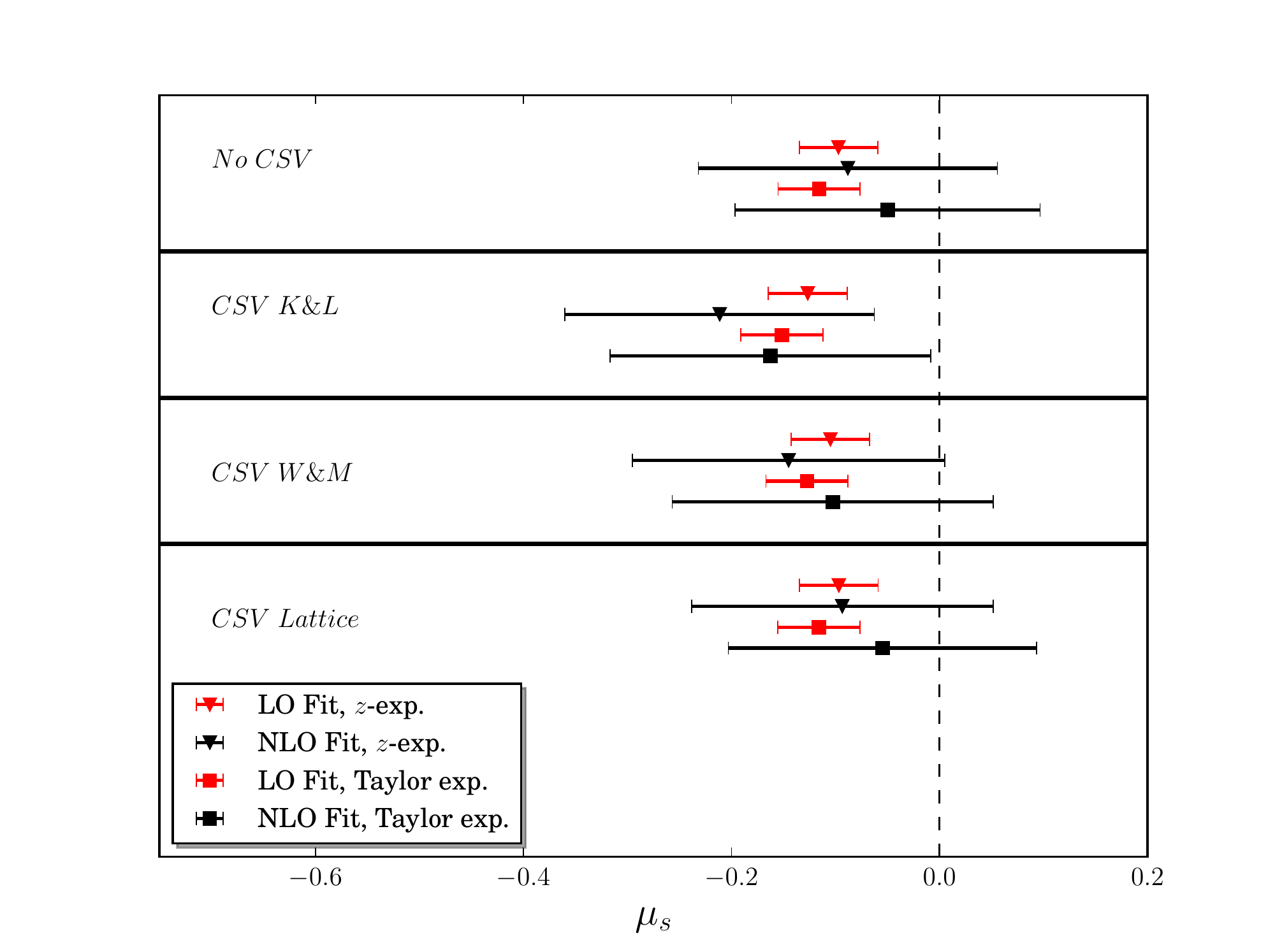}
\includegraphics[width=\columnwidth]{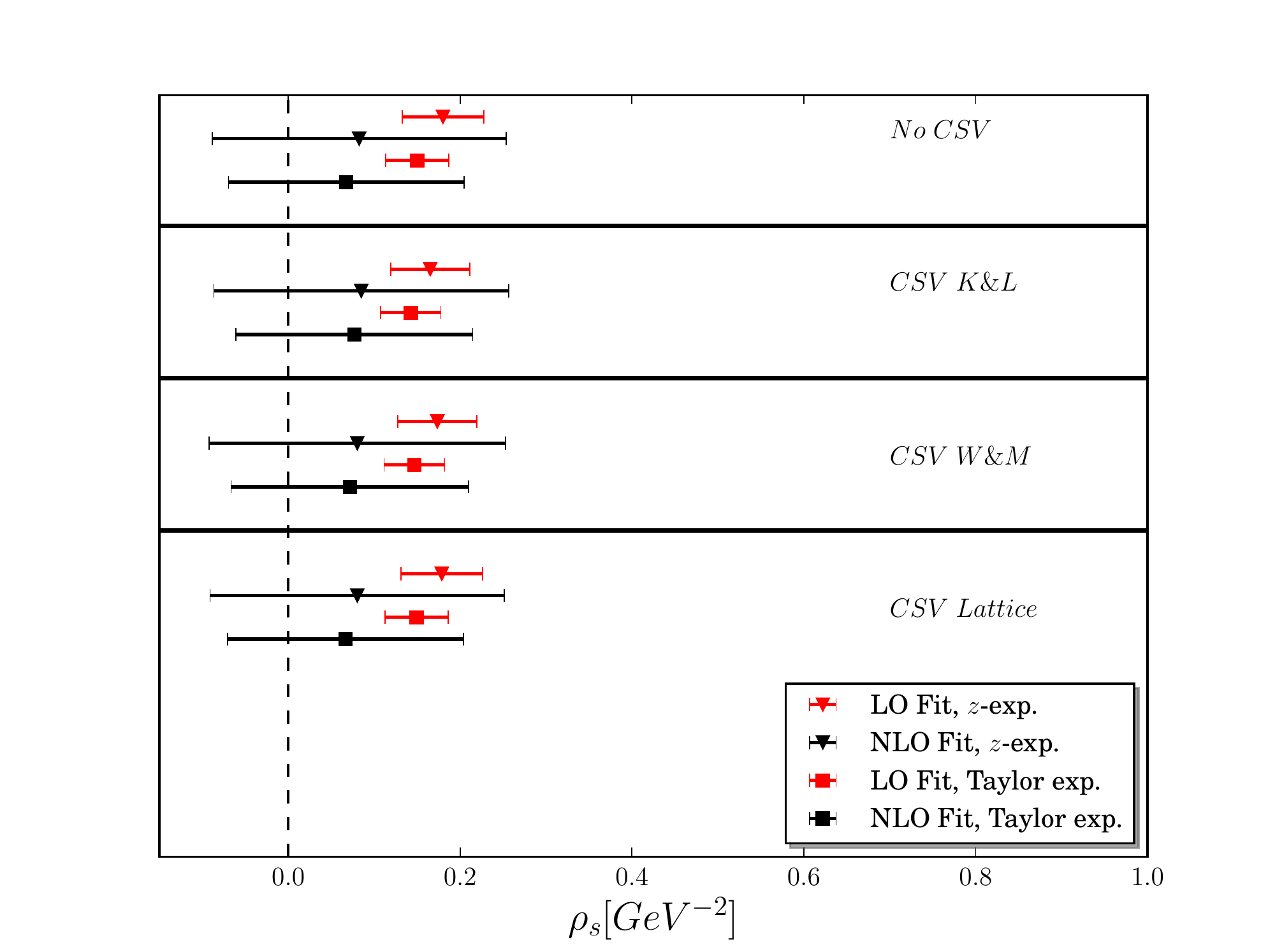}
\caption{Comparison  of determinations obtained from the present work
  with and without CSV  for the strange magnetic moment $\mu_s$ (left)
  and strange electric radius $\rho_s$(right).}
\label{Fig190}
\end{figure*}

In order to propagate the uncertainties, we extend
the covariance matrix above, Eq.~(\ref{eq31}), to include a correlated
uncertainty associated with the theoretical estimates of CSV.
For each theoretical description, we reanalyse the entire data set and
present in Fig.~\ref{Fig190} our determination of the strange magnetic
moment $\mu_s$ (left) and strange electric radius $\rho_s$ (right).
Since the Lattice CSV form factors are zero with a negligible
uncertainty, they are consistent with the ``No CSV'' results.
We also find no visible impact on $\mu_s$ and
$\rho_s$ from the inclusion of the W$\&$M CSV form factors.
Finally, when estimating the CSV form factors by the K$\&$L
parameters, we observe only small shifts in the central value of the
strangeness magnetic moment.
Nevertheless, even the ``worst case'' scenario of K$\&$L doesn't
appreciably affect the NLO fits.

\section{conclusion}
\label{sec5}
We have presented a complete global analysis of all PVES asymmetry
data for the proton, $^4$He and deuterium.
We have investigated the $\gamma Z$ exchange correction and the effect
that CSV form factors have on the extraction of strange quark
contribution.
Including $\gamma Z$ box contribution in the analysis leads to small
increases in the magnitude of the central values of $\mu_s$ and
$\rho_s$ when compared to results obtained without constraints from
$\gamma Z$-exchange.
CSV results considered in this work have tiny effects on the central
values of the strangeness parameters, with the largest effect, while
still small, coming from the inclusion of CSV form factors as provided
by Kubis and Lewis~\cite{18}.
Our results favour non-zero values for the strangeness magnetic moment
and electric radius, with the most significant constraints coming from
the LO fits using a Taylor expansion for the strangeness form factors.
Finally, in order to examine the model dependence of employing a
Taylor expansion in our analysis, we also included fits using the
$z$-expansion which were found to be in agreement.

The latest theory estimates on CSV are small --- indeed small enough
that they would not cloud the interpretation of future precision
strangeness measurements.
However, we note that the back-angle measurements do exhibit
sensitivity to the effective axial form factor, presenting an
opportunity for future investigation.
The combined efforts to improve the resolution of strangeness, and
reveal the structure of the anapole form factor offer the potential to
establish a precision era of QCD and the nucleon.
While further advancing the understanding of the mechanisms underlying
nonperturbative QCD, such work will serve to gain further confidence
in the use of lattice QCD for precision constraints in tests of the
standard model.

\acknowledgments

We would like to thank Wally Melnitchouk for useful discussions.
AA is supported by by Physics Department, Taif University.
RDY and JMZ are supported by the Australian Research Council under
grants FT120100821, FT100100005, DP140103067 and DP190100297.

\appendix

\section{Parity violating asymmetries}
Table \ref{tab1} lists the asymmetries and their dependence on the
leading unknown hadronic structure for all existing PVES experiments.

\begin{table*}
  \caption{\label{tab1}
    Values of $\eta_i$, appearing in Eq.~\ref{eq22}, which describe
    the theoretical asymmetry for each experiment (in
    partsper-million).
    $A^{phys}$ and $\delta A$ are the measured asymmetry and the
    corresponding uncertainty respectively where the statistic ans
    systematic error have been added in quadrature.
    While $\delta A_{cor}$ is the correlated error in the G0
    experiment~\cite{43,44}.}
\begin{ruledtabular}
\begin{filecontents*}{AsymmetryData36.txt}
%Exp. %Tar. %$Q^2$ %$\theta$ %E %$\eta_o$ %$\eta^p_A$ %$\eta^n_A$ %$\eta_E$  %$\eta_M$   %$A^{Phy}$  %$\delta_A$    %$\delta A_{Corr}$ %Ref
SAMPLE d      0.038    144    0.11   -2.13    0.46       -0.30   1.16    0.28    -3.51    0.81    0 \cite{35}         
SAMPLE d      0.091    144    0.18   -7.02    1.04       -0.65   1.63    0.77    -7.77    1.03    0 \cite{35}         
SAMPLE p      0.1      144    0.2    -5.50    1.57       0       2.11    3.45    -5.61    1.11    0 \cite{37}

HAPPEX p      0.477    12.3   3.35   -15.82   1.13       0       55.65   23      -15.05   1.13    0 \cite{42}    

PVA4 p        0.230    35.3   0.85   -5.78    0.88       0       22.45   5.08    -5.44    0.60    0 \cite{45}    
PVA4 p        0.108    35.4   0.57   -1.82    0.26       0       10.07   1.05    -1.36    0.32    0 \cite{147}     

G0 p          0.122    6.68   3.03   -1.93    0.06       0       11.94   1.17    -1.51    0.49    0.18 \cite{43}  
G0 p          0.128    6.84   3.03   -2.08    0.06       0       12.58   1.30    -0.97    0.46    0.17 \cite{43}  
G0 p          0.136    7.06   3.03   -2.28    0.07       0       13.43   1.47    -1.30    0.45    0.17 \cite{43}  
G0 p          0.144    7.27   3.03   -2.48    0.08       0       14.29   1.66    -2.71    0.47    0.18 \cite{43}  
G0 p          0.153    7.5    3.03   -2.73    0.09       0       15.27   1.89    -2.22    0.51    0.21 \cite{43}   
G0 p          0.164    7.77   3.03   -3.03    0.11       0       16.49   2.19    -2.88    0.54    0.23 \cite{43}  
G0 p          0.177    8.09   3.03   -3.41    0.13       0       17.93   2.58    -3.95    0.50    0.20 \cite{43}  
G0 p          0.192    8.43   3.03   -3.87    0.15       0       19.63   3.07    -3.85    0.53    0.19 \cite{43}  
G0 p          0.210    8.84   3.03   -4.45    0.19       0       21.69   3.72    -4.68    0.54    0.21 \cite{43}    
G0 p          0.232    9.31   3.03   -5.20    0.23       0       24.25   4.62    -5.27    0.59    0.23 \cite{43}    
G0 p          0.262    9.92   3.03   -6.29    0.31       0       27.82   6.03    -5.26    0.53    0.17 \cite{43}  
G0 p          0.299    10.63  3.03   -7.73    0.42       0       32.33   8.06    -7.72    0.80    0.35 \cite{43}    
G0 p          0.344    11.45  3.03   -9.61    0.58       0       37.97   11.01   -8.40    1.09    0.52 \cite{43}     
G0 p          0.410    12.59  3.03   -12.60   0.87       0       46.50   16.34   -10.25   1.11    0.55 \cite{43}  
G0 p          0.511    14.2   3.03   -17.61   1.49       0       60.11   27.01   -16.81   1.73    1.50 \cite{43}  
G0 p          0.631    15.98  3.03   -24.05   2.52       0       77.05   44.07   -19.96   1.69    1.31 \cite{43} 
G0 p          0.788    18.16  3.03   -33.08   4.45       0       100.26  74.55   -30.83   3.16    2.59 \cite{43}  
G0 p          0.997    20.9   3.03   -45.78   8.32       0       132.55  131.72  -37.93   11.55   0.52 \cite{43} 

HAPPEX He$^4$     0.091    6.0    2.91   -7.52 0         0       -20.22  0       -6.72    0.87    0 \cite{39}         
HAPPEX p          0.099    6.0    3.03   -1.41 0.04      0       9.54    0.76    -1.14    0.25    0 \cite{148}     

HAPPEX He$^4$     0.077    6.0    2.67   -6.37 0         0      -16.58   0       -6.40    0.26    0 \cite{149} 
HAPPEX p          0.109    6.0    3.18   -1.63 0.04      0       10.58   0.93    -1.58    0.13    0 \cite{149} 

PVA4 p          0.22     144.5  0.31 -13.31   3.47       0       2.88    11.13   -17.23   1.21    0 \cite{46} 

G0 p          0.221    110    0.35   -10.61   2.73       0       9.37    8.93    -11.25   0.9     0.43 \cite{44}   
G0 d          0.221    110    0.35   -15.25   2.05    -1.38      7.63    2.21    -16.93   0.91    0.21 \cite{44}
G0 p          0.628    110    0.68   -36.91   11.91      0       19.71   62.25   -45.9    2.53    1.0 \cite{44}  
G0 d          0.628    110    0.68   -50.72   8.46    -5.66      16.63   14.62   -55.5    3.86    0.7 \cite{44}  

HAPPEX p      0.624    13.7   3.48   -23.54   2.12       0       76.59   42.76   -23.8    0.86    0 \cite{150} 

PVA4 d        0.224    145    0.31   -18.65   2.50    -1.68      2.17    2.67    -20.11   1.35    0 \cite{189}

Qweak p          0.025    7.9   1.16   -0.22    0.006      0       2.27    0.05   -0.279    0.05   0 \cite{73}
\end{filecontents*}
\pgfplotstabletypeset[every head row/.style={before row=\toprule, 
                             after row=\hline},
   every last row/.style={after row=\bottomrule}, columns/4/.style={column type=c||},
columns/9/.style={column type=c||},
   col sep=space, 
   header=false,
    display columns/0/.style={column name={Experiment}}, 
    display columns/1/.style={column name={Target}},
    display columns/2/.style={column name={$Q^2$}},
    display columns/3/.style={column name={$\theta$}},
    display columns/4/.style={column name={$E$}},
    display columns/5/.style={column name={$\eta_0$}}, 
    display columns/6/.style={column name={$\eta^p_A$}},
    display columns/7/.style={column name={$\eta^n_A$}}, 
    display columns/8/.style={column name={$\eta_E$}},
    display columns/9/.style={column name={$\eta_M$}},
    display columns/10/.style={column name={$A^{phys}$}},
    display columns/11/.style={column name={$\delta A$}},
    display columns/12/.style={column name={$\delta A_{cor}$}},
   display columns/13/.style={column name={Ref.}},
    string type]{AsymmetryData36.txt}
\end{ruledtabular}
\end{table*}

\section{Energy dependence of the $\gamma$--$Z$ box}
\label{app:gZ}

To incorporate the effect induced by the energy-dependent component of
the $\gamma$--$Z$ box radiative correction, the measured PV
asymmetries given in Eq.~(\ref{eq10}) are modified by
\begin{equation}
  A^{p}_{PV\,corr} = A^{p}_{PV} -
  \left[
    \frac{-G_F Q^2}{4\sqrt{2}\pi\alpha}
    \right]
  \square_{\gamma Z}(E,Q^2)\ .
\end{equation}
The forward (or vanishing momentum transfer) limit of this box are
taken from Ref.~\cite{Hall:2015loa}, which extends Ref.~\cite{292} to
also incoporate duality constraints.
To estimate the momentum transfer dependence, we adopt the model
suggested by Gorchtein~{\em et al.} \cite{74}:
\begin{equation}
\square_{\gamma Z}(E,Q^2)=\square_{\gamma
  Z}(E,0)\frac{\exp\left(-BQ^2/2\right)}{F_1^p(Q^2)} \,,
\end{equation}
with slope parameter estimated to be $B=7\pm 1\gev^2$, and $F_1^p$ the
electromagnetic Dirac form factor of the proton.
The numerical values used in this work are summarised in
Table~\ref{tab100}.

\begin{table}
  \caption{\label{tab100}
    The $\square_{\gamma Z}
       (E)\ \big (\times 10^{-3}\big)$ corrections evaluated for the
    measured proton PV asymmetry $A^{p}_{PV}$ at forward angles.}
\begin{ruledtabular}
\begin{filecontents*}{GammaZBox.txt}
#Exp #Q^2(Gev^2) #E(GeV) #space #gZBox(E)(x10^{-3}) #gZBox(E,Q^2)(x10^{-3})
Qweak  0.025 1.165  5.120\ $\pm$\ 0.671
HAPPEx 0.099 3.030  7.205\ $\pm$\ 0.701
PVA4   0.108 0.570  2.843\ $\pm$\ 0.580
HAPPEx 0.109 3.180  7.250\ $\pm$\ 0.713
G0     0.122 3.030  6.969\ $\pm$\ 0.722
G0     0.128 3.030  6.907\ $\pm$\ 0.728
G0     0.136 3.030  6.825\ $\pm$\ 0.737
G0     0.144 3.030  6.742\ $\pm$\ 0.745
G0     0.153 3.030  6.648\ $\pm$\ 0.754
G0     0.164 3.030  6.534\ $\pm$\ 0.766
G0     0.177 3.030  6.398\ $\pm$\ 0.780
G0     0.192 3.030  6.243\ $\pm$\ 0.795
G0     0.210 3.030  6.056\ $\pm$\ 0.814
PVA4   0.230 0.850  3.257\ $\pm$\ 0.610
G0     0.232 3.030  5.831\ $\pm$\ 0.834
G0     0.262 3.030  5.527\ $\pm$\ 0.859
G0     0.299 3.030  5.161\ $\pm$\ 0.885
G0     0.344 3.030  4.733\ $\pm$\ 0.905
G0     0.410 3.030  4.143\ $\pm$\ 0.917
HAPPEx 0.477 3.350  3.749\ $\pm$\ 0.943
G0     0.511 3.030  3.340\ $\pm$\ 0.898
HAPPEx 0.624 3.480  2.740\ $\pm$\ 0.881
G0     0.631 3.030  2.547\ $\pm$\ 0.831
G0     0.788 3.030  1.753\ $\pm$\ 0.706
G0     0.997 3.030  1.038\ $\pm$\ 0.525
\end{filecontents*}
\pgfplotstabletypeset[every head row/.style={before row=\toprule, 
                             after row=\hline},
   every last row/.style={after row=\bottomrule},
   col sep=space, 
   header=false,
     display columns/0/.style={column name={Experiment}}, 
     display columns/1/.style={column name={$Q^2$ (GeV$^2$)}},
    display columns/2/.style={column name={$E$ (GeV)}},
   display columns/3/.style={column name={$\square_{\gamma Z}
       (E)\ \big (\times 10^{-3}\big)$}},
    string type]{GammaZBox.txt}
\end{ruledtabular}
\end{table}

\bibliographystyle{apsrev4-1}
\bibliography{RefbibNu,bibExtras}
\end{document}